\documentclass[journal]{IEEEtran}
\usepackage{amsmath,amsfonts}
\usepackage{algorithmic}
\usepackage{algorithm}
\usepackage{array}
\usepackage[caption=false,font=normalsize,labelfont=sf,textfont=sf]{subfig}
\usepackage{textcomp}
\usepackage{stfloats}
\usepackage{url}
\usepackage{verbatim}
\usepackage{graphicx}
\usepackage{cite}
\usepackage{caption}
\usepackage{float}
\usepackage{enumitem}
\usepackage{booktabs}
\usepackage{tabularx}
\usepackage{amsmath}
\usepackage{amssymb}
\usepackage{cleveref}
\usepackage{multirow}
\usepackage{placeins}
\usepackage{xcolor}
\begin{document}

\title{{\color{black}QSTAformer: A Quantum-Enhanced Transformer} for Robust Short-Term Voltage Stability Assessment against Adversarial Attacks}

\author{Yang Li, \textit{Senior Member, IEEE}, Chong Ma, Yuanzheng Li, \textit{Senior Member, IEEE}, Sen Li,

Yanbo Chen, \textit{Senior Member, IEEE}, Zhaoyang Dong, \textit{Fellow, IEEE}
\thanks{{Y. Li is with the School of Electrical Engineering, Northeast Electric Power University, Jilin 132012, China (e-mail: liyang@neepu.edu.cn).}

{C. Ma is with State Grid Shandong Electric Power Company Jiaozhou Power Supply Company, Jiaozhou 266300, China (email:machong58112233@163.com).}

{Y. Z. Li is with the School of Artificial Intelligence and Automation,
Huazhong University of Science and Technology, Wuhan 430074, China (email: Yuanzheng\_Li@hust.edu.cn).}

{Sen Li is with the Department of Civil and Environmental Engineering,
The Hong Kong University of Science and Technology, Hong Kong.
(e-mail: cesli@ust.hk).}

{Yanbo Chen is with the State Key Laboratory of Alternate Electrical Power System with Renewable Energy Sources, School of Electrical \& Electronic Engineering, North China Electric Power University, Beijing 102206, China (e-mail: chenyanbo@ncepu.edu.cn).}

{Z. Y. Dong is with the Department of Electrical Engineering, City University of Hong Kong, Kowloon, Hong Kong 
(e-mail: zydong@cityu.edu.hk).}

}}
\markboth{}
{Shell \MakeLowercase{\textit{et al.}}: A Sample Article Using IEEEtran.cls for IEEE Journals}

\maketitle

\begin{abstract}
Short-term voltage stability assessment (STVSA) is critical for secure power system operation. While classical machine learning-based methods have demonstrated strong performance, they still face challenges in robustness under adversarial conditions. This paper proposes {\color{black} QSTAformer—a tailored quantum-enhanced Transformer architecture} that embeds parameterized quantum circuits (PQCs) into attention mechanisms—for robust and efficient STVSA. A dedicated adversarial training strategy is developed to defend against both white-box and gray-box attacks.  Furthermore, diverse PQC architectures are benchmarked to explore trade-offs between expressiveness, convergence, and efficiency.  To the best of our knowledge, this is the first work to systematically investigate the adversarial vulnerability of quantum machine learning-based STVSA. Case studies on the IEEE 39-bus system demonstrate that {\color{black} QSTAformer achieves competitive accuracy, reduced complexity, and stronger robustness}, underscoring its potential for secure and scalable STVSA under adversarial conditions.
\end{abstract}


\section{Introduction}
\IEEEPARstart{W}{ith} the high penetration of converter-interfaced renewable energy sources and the growing deployment of fast-acting power electronic devices, maintaining short-term voltage stability (STVS) in modern power systems  has become a pressing challenge \cite{stankovic2021fault}. STVS characterizes a power system’s ability to preserve acceptable voltage profiles during the initial seconds following a disturbance \cite{hatziargyriou2020definition}, and this stability is primarily influenced by the dynamic behavior of fast acting loads, notably induction motors \cite{zhang2022load}. Given the high dimensionality and strong nonlinearity of STVS phenomena, there is a urgent need for methods that enable fast and accurate short-term voltage stability assessment (STVSA) \cite{2, zhu2023robust, li2024short}.

Currently, the primary approaches for STVSA in power systems include time-domain simulation \cite{lara2023revisiting}, energy function-based analysis \cite{3, kawabe2014analytical}, and data-driven methods. Time-domain simulation remains the most accurate technique by solving differential-algebraic equations in response to disturbances, while energy function-based methods offer faster yet approximate assessments based on system energy trajectories. However, the performance of both approaches heavily depends on the fidelity of the underlying system models and the accuracy of their parameters. {\color{black} Yet, the increasing integration of dynamic loads and converter-based renewables introduces significant uncertainty into system dynamics \cite{yang2021real, yang2024gaussian}.} This renders accurate modeling and parameter identification increasingly difficult \cite{6}, thereby motivating the need for alternative methods that can offer both adaptability and resilience under such complexity.

The widespread deployment of wide area measurement systems has enabled synchronized and high-precision data collection in power systems via phasor measurement units (PMUs) \cite{dasgupta2013real, guddanti2020pmu, yang2021real}. This development provides a foundation for implementing data-driven approaches to STVSA. In prior work, Li et al. \cite{li2023pmu} explored deep transfer learning strategies leveraging PMU-based time-series data to achieve accurate and adaptive STVSA under topological change conditions.
Reference \cite{mohammadniaei} proposed a distributed digital twin -based framework for multi-dimensional stability co-prediction in power systems. Reference \cite{huang2021resilient} designed a deep transfer learning approach based on a bi-directional Long Short-Term Memory (LSTM) network, aiming to  identify resilient network architectures with enhanced STVS performance.
Reference \cite{ren2019hybrid} introduced a time-adaptive intelligent system based on hybrid stochastic-integrated learning for STVSA.
Reference \cite{8} developed a heterogeneous edge-integrated graph attention network-based STVSA method utilizing a virtual homomorphism technique and a multilayer perceptron. As data imputation is a critical prerequisite in energy-system monitoring \cite{li2025ztfed}, Reference \cite{li2025multi} recently addressed STVSA under missing PMU data via a multi-task recurrent model for joint data imputation and stability assessment.

\textcolor{black}{
Delayed voltage recovery refers to the system’s inability to rapidly restore voltage levels after a disturbance—particularly when induction motor loads are involved. Reference \cite{zhang2018hierarchical} introduced a root-mean-square voltage dip severity index to quantify recovery performance across buses, enabling real-time STVS control. Reference \cite{xu2015assessing11} developed a hierarchical real-time assessment framework using a stochastic weighted neural network ensemble for dynamic security monitoring. Reference \cite{10856677} proposed a structure-based voltage stability index via dimensionality reduction to facilitate efficient assessment without requiring real-time data. Reference \cite{12} addressed data imbalance by combining data augmentation and Transformer architectures, achieving robust and accurate assessment under high renewable penetration.
}

While these machine learning methods have achieved significant successes, classical machine learning techniques continue to face challenges when managing large-scale datasets, high-dimensional feature spaces, and complex model architectures. These challenges include substantial demands for computational resources, \textcolor{black}{model convergence performance,} and specialized problem-solving capabilities. To overcome these obstacles, the adoption of quantum machine learning (QML) models presents a promising solution.

Compared with classical computation, quantum modeling offers theoretical exponential speedups in computing distances and inner products \cite{13}, enabling hybrid models to deliver comparable or superior performance with reduced computational overhead. Applications of quantum computing have already been explored in power system contexts. Applications of quantum computing have already been explored to some extent in power system contexts. For example, Zhou and Zhang \cite{zhou2022noise} proposed a quantum-based transient stability assessment method that employs a high-expressibility, low-depth circuit and quantum natural gradient descent, achieving improved  accuracy and noise resilience. Similarly, Ren et al. \cite{ren2024esqfl} developed a digital twin-driven quantum federated learning framework with explainability and privacy guarantees for decentralized voltage stability assessment.

Several recent efforts have aimed to enhance the adversarial robustness of machine learning models for STVSA. Li et al. \cite{li2025ai} proposed an LSTM-enhanced Graph Attention Network combined with Fast Gradient Sign Method (FGSM)-based adversarial training to improve model resilience against composite cyber-attacks. Ren et al. \cite{ren2021vulnerability} conducted a comprehensive vulnerability analysis of machine learning-based models under white-box and black-box attack scenarios and proposed mitigation strategies to enhance robustness. In parallel, reference \cite{23} explored the credibility of adversarial examples and their impacts on voltage stability evaluations in learning-based settings. Furthermore, Li et al. \cite{li2025enhancing} developed a deep reinforcement learning-based dispatch method with adversarial training to improve the cyber-resilience of integrated energy systems. These studies reveal the growing interest in adversarially robust models in power system applications.

Nevertheless, most existing approaches rely on classical machine learning architectures, which are constrained by high computational costs, limited model convergence capabilities, and difficulties in generalizing to complex dynamic systems. More importantly, current adversarial defense strategies remain primarily focused on improving robustness under classical settings, without considering how emerging quantum models may behave under similar threat scenarios. Although QML models offer theoretical advantages in scalability and expressive power, their security vulnerabilities—particularly under adversarial conditions—have not been systematically investigated in power system contexts. To fill this gap, this study develops QSTAformer—a task-tailored quantum-enhanced Transformer architecture for STVSA—by integrating parameterized quantum circuits (PQCs) into the attention mechanism, and systematically investigates its adversarial robustness through a dedicated training strategy. {\color{black}In addition, we incorporate two supporting modules from our previous work \cite{li2022deep}: a Semi-Supervised Fuzzy C-Means (SFCM) algorithm for soft labeling and a Least Squares Generative Adversarial Network (LSGAN)-based method for data augmentation. These components enrich and balance the training dataset, thereby enhancing  the generalization and robustness of the proposed classification model.}

To this end, we construct a unified STVSA learning framework that integrates semi-supervised labeling (via SFCM), data augmentation (via LSGAN), and quantum-enhanced classification (via QSTAformer). The core contributions of this paper are summarized as follows:
\begin{enumerate}
\item \textbf{Quantum-enhanced Transformer architecture:}  
    We develop {\color{black} QSTAformer, a tailored Transformer model for STVSA} that integrates PQCs into its self-attention mechanism. This design facilitates expressive modeling of nonlinear and high-dimensional power system dynamics while maintaining a lightweight and parameter-efficient structure.

    \item \textbf{Optimization and benchmarking of quantum circuit structures:}  
    Diverse PQC configurations are benchmarked with respect to training convergence, representation power, and inference efficiency. These insights guide circuit-level architecture design for scalable and efficient deployment of quantum-enhanced models in STVSA.

    \item \textbf{Adversarial robustness evaluation and defense strategy:}  
    {\color{black}To the best of our knowledge, this is the first study that systematically explores the adversarial vulnerability of QML models in STVSA} and proposes a tailored adversarial training mechanism to enhance resilience against both white-box and gray-box attacks.

    \item \textbf{Validation of secure and scalable STVSA under adversarial attacks:}  
    Experimental results on the IEEE 39-bus power system demonstrate that the QSTAformer model achieves competitive accuracy, reduced complexity, and stronger robustness compared to classical baselines, validating its potential as a secure and scalable solution for QML-enabled STVSA under adversarial conditions.
\end{enumerate}

\section{Background And Preliminaries}
This section presents the essential background concepts and methodological components that underpin the proposed STVSA framework. We first review the basic principles of quantum computing, which provide the foundation for the QSTAformer architecture. Next, we introduce key adversarial attack mechanisms relevant to power system security. Finally, we describe two auxiliary modules integrated into our framework: a SFCM algorithm for soft labeling and a LSGAN-based strategy for data augmentation.
\subsection{Basic Principles of Quantum Computing }
Quantum computing's core characteristics lie in its superposition and entanglement properties. The quantum superposition nature allows quantum bits to be in different states at the same time, while quantum entanglement establishes strong connections between quantum bits, and effective data encoding enhances the model's expressiveness and data handling capabilities \cite{21.5}. 
Quantum computing maps data from Euclidean space, which describes classical physical phenomena, to the infinite-dimensional Hilbert complex space. Using the superposition and entanglement properties of quantum states, enabling parallel and high-speed data computation, provides a novel approach to large-scale data processing.

A quantum bit (qubit), the fundamental unit of quantum computing, can be represented as a superposition of the two basis states, $|0\rangle$ and $|1\rangle$:
\begin{equation}
    \mid\psi\rangle=\alpha\mid0\rangle+\beta\mid1\rangle
\label{eq:6}
\end{equation}
Here, $\alpha$ and $\beta$ are complex coefficients and satisfy the normalization condition 
$|\alpha|^2+$$|\beta|^2=1.$

For system of $n$ quantum bits, the quantum state $\left|\psi\right\rangle$ can be expressed as:
\IEEEpubidadjcol
\begin{equation}
\begin{aligned}
|\psi\rangle = \sum_{i=0}^{2^n-1} \alpha_i |i\rangle = \bigotimes_{k=0}^{n-1} \left( \sum_{i_k=0}^{1} \alpha_{i_k} |i_k\rangle \right)
\label{eq:7}
\end{aligned}
\end{equation}
where each complex coefficient satisfies the normalization condition:
$\sum_{i=0}^{2^n-1}|\alpha_i|^2=1$. That is, the states of the $n$ quantum bits correspond to the $2^n$ state vectors in an $n$- 
dimensional Hilbert space, and the space consists of the tensor product of $n$ two dimensional 
Hilbert spaces\cite{23}.

Quantum entanglement describes the correlated state among multiple quantum systems and, for two quantum bits, can be represented as a tensor product:
\begin{equation}
\begin{aligned}
\mid \Phi^+ \rangle &= \frac{1}{\sqrt{2}}(\mid 0 \rangle \otimes \mid 0 \rangle + \mid 1 \rangle \otimes \mid 1 \rangle)
\label{eq:8}
\end{aligned}
\end{equation}

In fact, a quantum circuit consists of a series of quantum gates arranged to perform a specific algorithm or computational task through the manipulation of quantum bits. Different quantum gates 
have different functions, taking the example of a rotating X-gate as follows:
\begin{equation}
R_x(\theta) = \begin{bmatrix}
\cos(\theta/2) & -i\sin(\theta/2) \\
-i\sin(\theta/2) & \cos(\theta/2)
\label{eq:9}
\end{bmatrix}
\end{equation}
where ~\eqref{eq:9} represents the unitary operation of a single quantum bit rotated by an angle $\theta$ around the X-axis of the Bloch sphere. The effect is to rotate the quantum state in the X-axis direction by $\theta$, changing the superposition phase relation of the state vector.
Main diagonal elements reflecting the basis preservation property of the rotation, and the sub-diagonal element introduces imaginary units in response to the complex phase change of the rotation.
When the angle is given, it can be expressed as:
\begin{equation}
R_x(\theta)=e^{-i\theta\\X/2}
\label{eq:10}
\end{equation}
where $X$ denotes the bubbleley matrix, $i$ denotes imaginary units.

Quantum observation, a crucial step in quantum computing, involves measuring a quantum system. This measurement collapses the system from a superposition state to a definite ground state, determining the output of quantum information processing. It underpins applications like quantum computation, encryption, and communication.

To construct quantum circuits, the concept of parameter optimization in classical machine 
learning is borrowed. By using updatable parameter quantum gates such as rotating gates, $R_x(\theta),R_y(\theta),R_z(\theta)$ etc. to construct circuits, the variable parameters can be made to be continuously optimized to minimize or maximize the objective function for a particular task during the training process.
Approximating the model objective function by utilizing the accelerated computation capabilities of parameterized quantum circuits (PQCs)\cite{22}.
With the parameter optimization process of classical machine learning, which constantly uses PQCs to train the mapping between inputs and outputs, is a key part of QML. 
\subsection{Adversarial Attacks }
The core idea behind adversarial attacks is to exploit gradient information to mislead machine learning model predictions through carefully crafted input perturbations \cite{ren2021vulnerability, zhao2023robust, li2025ai}. Depending on the attacker's knowledge of the target model, adversarial attacks are typically categorized into three types: white-box, gray-box, and black-box attacks \cite{22.5}.

A white-box attack is an adversarial scenario where the attacker possesses full knowledge of the target model, covering its architecture, parameters, and training data. This scenario is widely studied in academia as it is the most in-depth attack on the model.
A more pertinent example is the gray-box attack scenario, in which the attacker has limited access to certain information, such as the type of input data and the probability of the output result, and where the basic principle is to iteratively construct and optimize adversarial samples by feeding back on the output of the model. Finally, in black-box attack scenarios, it is generally assumed that no information is known about the interior of the model, and the behavior of the model can only be understood through a large number of queries, which is costly and can be easily stopped by simple defensive measures. In this study, we focus on white-box and gray-box attack scenarios. {\color{black} Although black-box attacks are relevant in certain contexts, they are not considered in this study due to their limited threat level in power systems. These attacks typically rely on extensive querying and offer minimal model insight, making them inefficient and more easily detected in critical infrastructure environments. In contrast, gray-box attacks strike a practical balance between full and zero knowledge assumptions, effectively representing adversarial behaviors in cyber-physical systems where partial access or sensor feedback is commonly available.}

The following is an explanation of the principle of adversarial attack. 
The perturbation is defined as: 
\begin{equation}
    x'=x+\epsilon\quad\|\epsilon\|_p<\epsilon_{max}
\label{eq:11}
\end{equation}
where $x$ represents the unperturbed original input, and $\epsilon$ signifies a small perturbation introduced to this input. 
The $p$ norm of $\epsilon$ should remain within the maximum allowable range to ensure the perturbation remains imperceptible.

As the data of the model is tampered with, it can cause the classification results of the model to deviate from the true results. The connection between the perturbed input $x^{\prime}$ and the resulting predicted output  $y^{\prime}$
is described as follows: 
\begin{equation} 
y'=f(x')=\phi(W^Tx'+b)=\phi(W^T(x+\epsilon)+b)\quad\
\label{eq:12}
\end{equation}
where $y^{\prime} $ denotes the predicted output of the model, given the perturbed input $x^\prime$, $\phi$ represents 
the activation function used by the model, and $W$ and $b$ stand for the weights and biases of the model, respectively. $\epsilon$ represents a small perturbation.

As can be seen from \eqref{eq:12},  $W^Tx+b$ is the output obtained 
through the model from the original input $x$, and $W^T\epsilon$ is the effect of the perturbation. 
Note that even if $\epsilon$ is small, this small change may cause the output $y'$ to be significantly different from the original output $y$ due to the nonlinear nature of the activation function $\phi$.

This paper investigates three adversarial attack methods, including the Momentum Iterative Fast Gradient Sign Method (MI-FGSM), Projected Gradient Descent (PGD)  \cite{24}\cite{25}, which are both iteration-based adversarial attack methods, and Carlini \& Wagner (C\&W) Attack. MI-FGSM introduces a momentum term to enhance attack stability and effectively evades model defenses by accumulating gradient information at each iteration step, while PGD ensures attack concealment by projecting adversarial samples into an $\epsilon$-sphere, thereby maintaining the allowable perturbation range and generating destructive samples.
Unlike the two, the C\&W Attack is an optimization-oriented approach to adversarial attacks by constructing a loss function that allows the adversarial samples to be minimally perturbed while maintaining adversariality, and is also considered to be one of the most powerful adversarial attacks currently available \cite{26}.
\subsection{Semi-Supervised Fuzzy C-Means}
Despite the critical importance of short-term voltage stability assessment (STVSA), there remains a lack of universally accepted quantitative criteria for reliably distinguishing stable from unstable operating conditions. This complicates the assignment of ground-truth labels in large-scale datasets. As pointed out by Zhu et al. \cite{27}, manually labeling samples is time-consuming and unsuitable for real-time applications. Nevertheless, certain cases—such as persistent voltage drops below 0.7 p.u. or voltages consistently above 0.9 p.u.—can be confidently labeled using domain knowledge, providing a basis for semi-supervised learning.

{\color{black} This partial labeling insight motivates the adoption of semi-supervised clustering methods to effectively handle large volumes of unlabeled data. To this end, we employ a Semi-Supervised Fuzzy C-Means (SFCM) algorithm to infer soft labels for unlabeled samples. This process enriches the training dataset and enhances the generalization ability of the downstream classification model \cite{27}.}

SFCM is a semi-supervised fuzzy clustering algorithm that combines partially labeled and unlabeled data to guide the clustering process. By incorporating available label information into the objective function, the algorithm improves clustering accuracy and robustness \cite{12}. This clustering strategy aligns with our earlier work on hybrid data preprocessing for STVSA \cite{li2022deep}, where clustering and data augmentation were jointly employed to enhance feature separability in STVSA tasks. Its objective function is described as:

\begin{equation}
\begin{aligned}
\min J &= \sum_{i=1}^m \sum_{j=1}^n u_{ij}^2 \left\| x_i - c_j \right\|^2 \\
&\quad + \lambda \sum_{i=1}^m \sum_{j=1}^n \left(u_{ij} - f_{ij} b_j\right)^2 \left\| x_i - c_j \right\|^2
\label{eq:sfcm}        
\end{aligned}
\end{equation}
where $n$ represents the total number of data points, $i$ is the index for data points, $m$ indicates the total number of cluster centers, $j$ is the index for cluster centers. $u_{ij}$ represents the membership degree of data point $x_i$ to cluster center $c_j$, $f_{ij}$ indicates whether there is prior knowledge about the relationship between data point $x_i$ and cluster center $c_j$, and $b_j$ is the expected characteristic of the cluster center defined by prior knowledge. $\lambda$ is a regularization parameter used to balance the influences of unsupervised and semi-supervised learning.

The SFCM adjusts the calculation of affiliation by introducing external knowledge, where the affiliation formula is expressed as:

\begin{equation}
\begin{aligned}
u_{ij}=\frac{1}{1+\lambda}\left[\frac{1+\lambda\left(1-b_j\sum_{i=1}^mf_{ij}\right)}{\sum_{k=1}^c\left(\frac{\left\|x_i-c_j\right\|}{\left\|x_i-c_k\right\|}\right)^2}+\lambda f_{ij}b_j\right]
\label{eq:uij}
\end{aligned}
\end{equation}

The update formula for the center of mass is given by:  
\begin{equation}
\begin{aligned}
c_j=\frac{\sum_{i=1}^mu_{ij}^mx_i}{\sum_{i=1}^mu_{ij}^m}
\label{eq:vj}
\end{aligned}
\end{equation}
through the continuous updating of affiliation and center of mass, it is determined whether the change in the cluster center is less than a set 
threshold; if yes, the algorithm converges, otherwise the iterative updating is repeated.

\subsection{Least Squares Generative Adversarial Networks}
{\color{black} To improve the diversity and representativeness of the training dataset, we employ an LSGAN-based data augmentation strategy. In STVSA tasks, rare or borderline scenarios are often underrepresented, potentially leading to biased or overfitted models. By synthesizing realistic voltage trajectories conditioned on cluster-informed labels, LSGAN enriches sparse regions in the feature space and complements the semi-supervised clustering, thus enhancing the generalization capacity of the downstream quantum-enhanced classifier.}

LSGAN, introduced by Mao et al. in 2017 \cite{mao2017least}, is a variant of the traditional Generative Adversarial Network (GAN). Its core improvement lies in replacing the standard cross-entropy loss with a least squares loss, which stabilizes training and mitigates gradient vanishing. This enhancement contributes to improved convergence and higher quality of generated data. The objective function of LSGAN is formulated as \cite{li2021privacy}:
\begin{subequations}
\renewcommand{\theequation}{\theparentequation\alph{equation}}  

\begin{equation}
\begin{aligned}
    \min_{D} V_{\text{LSGAN}}(D) &= \frac{1}{2} \mathbb{E}_{x \sim p_{\text{data}}(x)} \left[(D(x) -1)^2\right] \\
    &\quad + \frac{1}{2} \mathbb{E}_{z \sim p_z(z)} \left[D(G(z)))^2\right]
\end{aligned}
\label{eq:minD}
\end{equation}

\begin{equation}
\begin{aligned}
    \min_G V_{\text{LSGAN}}(G) &= \frac{1}{2} \mathbb{E}_{z \sim p_z(z)} \left[(D(G(z)) - 1)^2\right]
\end{aligned}
\label{eq:minG}
\end{equation}
\end{subequations}
where equation \eqref{eq:minD} and \eqref{eq:minG} represent the discriminator $D$ and generator $G$ objective functions, where the former wants to recognize as much false data as possible, while the latter wants to deceive the discriminator as much as possible.  $x$ denotes a sample of real data, $z$ is the potential spatial noise vector that serves as input to the generator, $\mathbb{E}$ denotes the expected value operator.  $p_{\mathrm{data}}(x)$ and $p_{\mathrm{z}}(z)$
denote the probability distributions of the real dataset and some kind of noisy dataset, respectively.
The notation $D(G(z))$ denotes the discriminative output score assigned by the discriminator $D$ to the generated data $G(z)$, where $G(z)$ is the synthetic sample produced by the generator $G$ from the latent noise vector $z$,
$D(x)$ indicates the probability that the discriminator classifies the input data $x$ as real.

{\color{black}
\section {QSTAformer: A Quantum-Enhanced Transformer Architecture for STVSA}
Inspired by recent advances in quantum self-attention mechanisms such as the Quantum Adaptive Transformer \cite{chen2025quantum}, we propose QSTAformer: a hybrid quantum-classical Transformer architecture specifically tailored for STVSA. In contrast to prior general-purpose quantum Transformers, QSTAformer is designed with domain-specific enhancements: it integrates PQCs directly into the attention mechanism and incorporates adversarial training as well as semi-supervised and generative data augmentation to meet the unique challenges of cyber-physical power systems. The following subsections elaborate on the QSTAformer’s architectural design and its embedded PQC for expressive quantum representation within the attention mechanism.

\subsection {Quantum Enhanced Transformer Model Principles}
Transformer is a neural network architecture based on a self-attention mechanism, whose core advantage lies in its ability to process sequential data in parallel and effectively model long-range dependencies. The traditional Transformer's self-attention mechanism calculates the attention weights through the Query (Q), Key (K) and Value (V) matrices. However, this classical architecture faces the challenge of $O(n^2)$ computational complexity when dealing with complex time-series data.

In this paper, we adopt the QSTAformer architecture to realize hybrid quantum-classical sequence modeling by integrating PQCs in the last coding layer of the traditional Transformer, replacing part of the classical attention module. The design adopts a progressive architecture of \emph{N}-1 classical Transformer coding layers plus one quantum-enhanced coding layer, with the classical layer in the early stage ensuring stable gradient propagation, and the final quantum layer providing non-classical feature transformation capability.

The architecture of the QSTAformer model is depicted  in Fig.\ref{QSTAformer-1}.The function of each structure in the proposed QSTAformer model can be expressed by mathematical formulations.
\begin{figure}[htbp]
    \centering
    \includegraphics[width=1\linewidth]{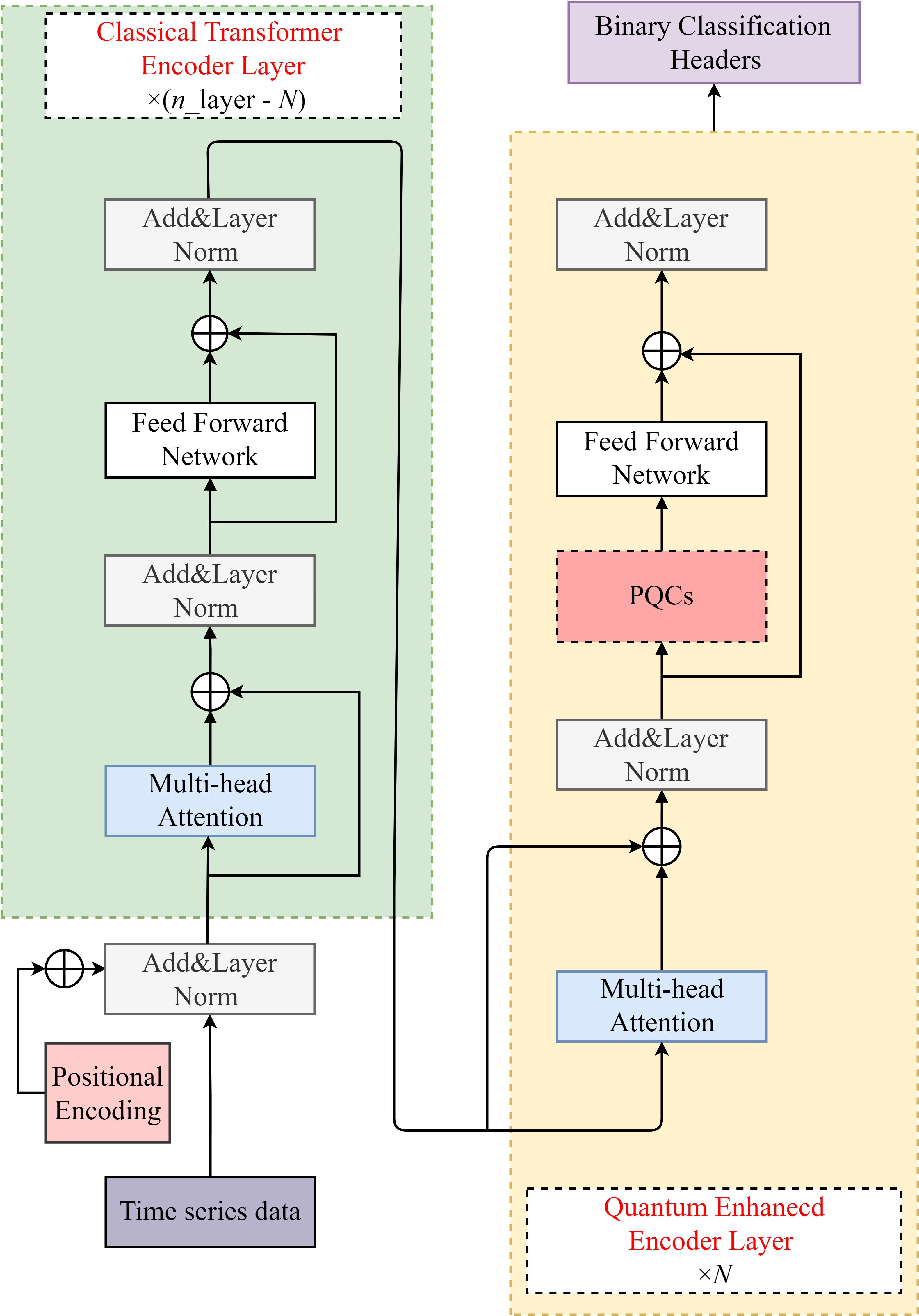}
    \caption{\textcolor{black}{Architecture for the QSTAformer}}
    \label{QSTAformer-1}
\end{figure}

\paragraph{Input Embedding Layer}
\begin{equation}
h_0 = \mathrm{LayerNorm}(W_e x + b_e), \quad h_0 \in \mathbb{R}^{L \times d}
    \label{eq:Input}
\end{equation}
    where $x \in \mathbb{R}^{L \times 1}$ is the original input  $W_e \in \mathbb{R}^{d \times 1}$ is the embedding weight matrix, $b_e \in \mathbb{R}^{d}$ is the embedding bias vector, $L$ is the sequence length, and $d$ is the hidden dimension. The input embedding layer projects one-dimensional time series data into a high-dimensional feature space, providing rich representational capability for subsequent processing.

\paragraph{Positional Encoding Layer}
\begin{equation}
h_0 \leftarrow h_0 + PE
    \label{eq:Positional}
\end{equation}
    where $PE \in \mathbb{R}^{L \times d}$ is the sinusoidal positional encoding matrix. This layer injects temporal information for each position, enabling the model to understand the relative and absolute positions of tokens in the sequence.

\paragraph{Classical Transformer Encoder Layers}
\begin{equation}
z_i = \mathrm{LayerNorm}(h_{i-1} + \mathrm{MultiHeadSelfAttention}(h_{i-1}))
    \label{eq:Transformer Encoder}
\end{equation}

\begin{equation}
h_i = \mathrm{LayerNorm}(z_i + \mathrm{FFN}(z_i)), \quad i = 1, 2, \ldots, N-1
    \label{eq:FFN}
\end{equation}

where the Multi-Head Self-Attention captures long-range dependencies, and the feedforward network (FFN) applies position-wise non-linear transformations to enhance representation learning. The FFN is defined as:

\begin{equation}
\mathrm{FFN}(x) = \max(0, xW_1 + b_1)W_2 + b_2
    \label{eq:FFN_def}
\end{equation}
where $W_1 \in \mathbb{R}^{d \times d_f}$, $W_2 \in \mathbb{R}^{d_f \times d}$ are weight matrices, and $b_1, b_2 \in \mathbb{R}^d$ are bias vectors.

\paragraph{ Quantum Encoder Layer}

The Quantum Encoder Layer enhances feature extraction by incorporating quantum operations into the classical Transformer structure, enabling the capture of non-classical correlations and improving model expressiveness.

Initially, classical multi-head attention and layer normalization are applied to extract contextual features (omitted here for brevity to focus on the quantum design).

Next, each time-step feature is projected into a quantum-compatible low-dimensional space using a learnable quantum projection matrix:

\begin{equation}
h_q = \tanh(W_q' h'_i) + t
\label{eq:tanh}
\end{equation}
where \(W_q' \in \mathbb{R}^{n \times d}\) is a learnable quantum projection matrix matching the qubit count \(n\), \(t\) is the time-step conditional signal, and the \(\tanh\) activation function ensures values remain within valid ranges for quantum gate parameters. This step compresses high-dimensional features while preserving key information for quantum circuit processing.

The projected features are then processed by a PQC for non-classical feature transformation and entanglement-based correlation extraction:
\begin{equation}
QC(h_q) = [\langle Z_j \rangle ]_{j=1}^{n}
    \label{eq:QC(h_q)}
\end{equation}
where \(QC(\cdot)\) represents the PQCs outputs as the expectation values of Pauli-Z measurements, enabling feature representation capabilities beyond classical methods.

The quantum measurement results are reintegrated into the original feature pathway via linear projection and residual connection:
\begin{equation}
z_i = h'_i + W_o \cdot QC(h_q)
    \label{eq:z_i}
\end{equation}
where \(W_o \in \mathbb{R}^{d \times n}\) is the output projection matrix mapping quantum measurements back into the classical feature space while maintaining gradient flow stability, aiding training convergence.

Finally, a feedforward network with Gaussian error linear unit (GELU) activation further refines the fused features to output the quantum-enhanced encoded result:

\begin{equation}
h_N = \mathrm{LayerNorm}(z + \mathrm{FFN}(z))
    \label{eq:h_N}
\end{equation}
Through this quantum-classical hybrid design, the Quantum Encoder Layer leverages quantum superposition and entanglement to capture complex high-dimensional correlations that traditional Transformers cannot, while maintaining stability and scalability. This design effectively enhances representational capability for sequence modeling and classification tasks, introducing a new quantum-enhanced processing dimension to deep learning with significant theoretical and practical value.

\paragraph{Output Prediction Layer}

For sequence-level classification tasks, QSTAformer extracts the global representation of the final time step for classification:

\begin{equation}
\mathrm{logits} = W_{cls} \cdot h_N[L] + b_{cls}
    \label{eq:logits}
\end{equation}
where $h_N[L] \in \mathbb{R}^d$ denotes the last time step feature of the quantum encoder output sequence, carrying the global sequence representation enhanced by quantum processing. Here, $W_{cls} \in \mathbb{R}^{c \times d}$ is the classification weight matrix, and $b_{cls} \in \mathbb{R}^c$ is the classification bias vector, with $c$ representing the number of target classes. This linear transformation projects high-dimensional sequence features into the class space, generating logits for each class.

The predicted probabilities are computed by:

\begin{equation}
\hat{y} = \mathrm{softmax}(\mathrm{logits})
    \label{eq:softmax}
\end{equation}

The $\mathrm{softmax}$ function converts raw scores into a probability distribution, ensuring $\sum_{i=1}^{c} \hat{y}_i = 1$ with $\hat{y}_i \in [0, 1]$. The final predicted class is determined using the maximum probability criterion.This design fully utilizes the rich feature representations provided by the quantum encoder layer, particularly the non-classical feature correlations captured through quantum entanglement, thereby enhancing the discriminative capability for complex classification tasks.

The QSTAformer model achieves effective fusion of quantum computing and deep learning. Through quantum superposition and entanglement mechanisms, this model gains exponential-level feature representation capability in quantum Hilbert space, breaking through the expressive limitations of classical Transformers. The hybrid architecture design ensures training stability through classical layers while providing non-classical feature transformations through quantum layers, effectively solving the gradient vanishing problem of pure quantum models and reducing parameter complexity. This introduces a new processing dimension for traditional neural networks, enabling QSTAformer to demonstrate potential beyond classical models in both theoretical and practical applications.

\subsection{Variational Quantum Circuit Structure}
At the heart of the QSTAformer model lies the use of PQC to augment the attention module in the classical Transformer coding layer, thereby hopefully fully mining and extracting features and correlations between features that have been mapped to the compressed data in Hilbert space. 
The PQC adopts a layered architecture, with each layer containing four core components: data encoding, parameterized variation, quantum entanglement, and quantum measurement to realize complex quantum feature transformations. The generic structure of the PQC components used in this paper is shown in Fig. \ref{PQC}.In the design of quantum layers, each layer of quantum circuits consists of the following main stages:

\begin{figure}[htbp]
    \centering
    \includegraphics[width=\linewidth]{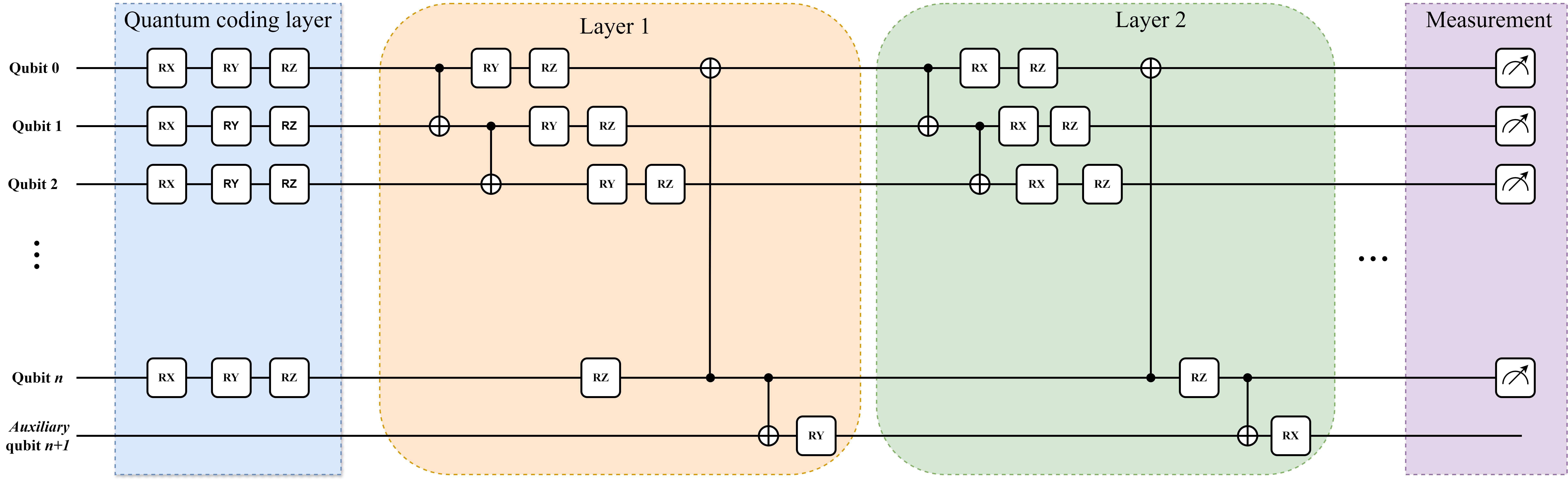}
    \caption{\textcolor{black}{Generalized structure of the applied ring entanglement connection}}
    \label{PQC}
\end{figure}

\paragraph{Data Encoding Layer}
The process begins by encoding classical input data into quantum states through the combined use of single-bit gates. These gates rotate the state of each bit according to the classical input features so that the quantum state represents the classical data in a high-dimensional quantum Hilbert space (in the case of the X gate).
\begin{equation}
\mathrm{RX}(\theta) = e^{-i \theta X / 2}
    \label{eq:Input2}
\end{equation}
    where \( X \) is the Pauli-X matrix, and \( \theta \) is the rotation angle.The collaborative rotation strategy between different gates ensures that the characteristic information of classical data can be efficiently encoded into quantum states, providing a rich initial state for subsequent quantum transformations.

\paragraph{Variational Quantum Layer}
The variational quantum layers utilize learnable, parameterized quantum gates to transform quantum states in accordance with the task objectives, thereby enabling the system to acquire optimal feature representations through quantum operations. These gates dynamically modulate the quantum states, allowing the circuit to learn and encode meaningful patterns from the data. Functioning as the quantum counterparts of trainable weights in classical neural networks, these parameterized gates facilitate expressive and non-linear feature transformations within the quantum state space. This incorporation of trainability into quantum circuits enhances the model’s adaptability across diverse tasks while capitalizing on the exponential representational capacity inherent to quantum systems.

\paragraph{Quantum Entanglement Layer}

\begin{equation}
\mathrm{CNOT}(i \rightarrow (i+1)\bmod n), \quad \forall i \in \{0, \ldots, n-1\}
    \label{eq:Entanglement1}
\end{equation}

\begin{equation}
\mathrm{CNOT}(n-1 \rightarrow n), \quad R_Y(\theta_{l,n}) \; \text{on auxiliary qubit } n
    \label{eq:Entanglement2}
\end{equation}
where CNOT represents controlled-NOT gates. The operation 
\(\mathrm{CNOT}(i \rightarrow (i+1)\bmod n)\) 
implements ring topology entanglement connections, where the modulo operation 
\(\bmod n\) 
ensures the \(n-1\)-th qubit connects with the \(0\)-th qubit, forming a complete ring structure. 
This enables all qubits to mutually influence each other through entanglement, capturing global correlations between features. 
Meanwhile, the introduction of auxiliary qubit \(n\) implements additional control capability. 
\(\mathrm{CNOT}(n-1 \rightarrow n)\) entangles the last main qubit with the auxiliary qubit \(n\), and 
\(R_Y(\theta_{l,n})\) applies parameterized rotation on the auxiliary qubit. 
The auxiliary qubit serves as a convergence point for global information, capable of capturing collective behavior of the entire quantum system, enhancing the circuit's expressive capability.

\paragraph{Quantum Measurement Layer}
\begin{equation}
QC(h_q) = [\langle Z_j \rangle ]_{j=1}^{n},
    \label{eq:Measurement}
\end{equation}
    where \(\langle Z_j \rangle = \langle \psi | Z_j | \psi \rangle\). Each qubit is measured using the Pauli-Z observable, returning the expectation value for each qubit, which is then used for subsequent classical computation such as classification or regression tasks. This process translates the quantum-enhanced representations into classical values while preserving the non-classical correlations established by the quantum circuit, enabling the seamless integration of quantum processing advantages into hybrid quantum-classical models for real-world applications.

}
\section{Proposed Methodology}

\begin{figure*}[htbp]  
    \centering
    \includegraphics[width=\textwidth]{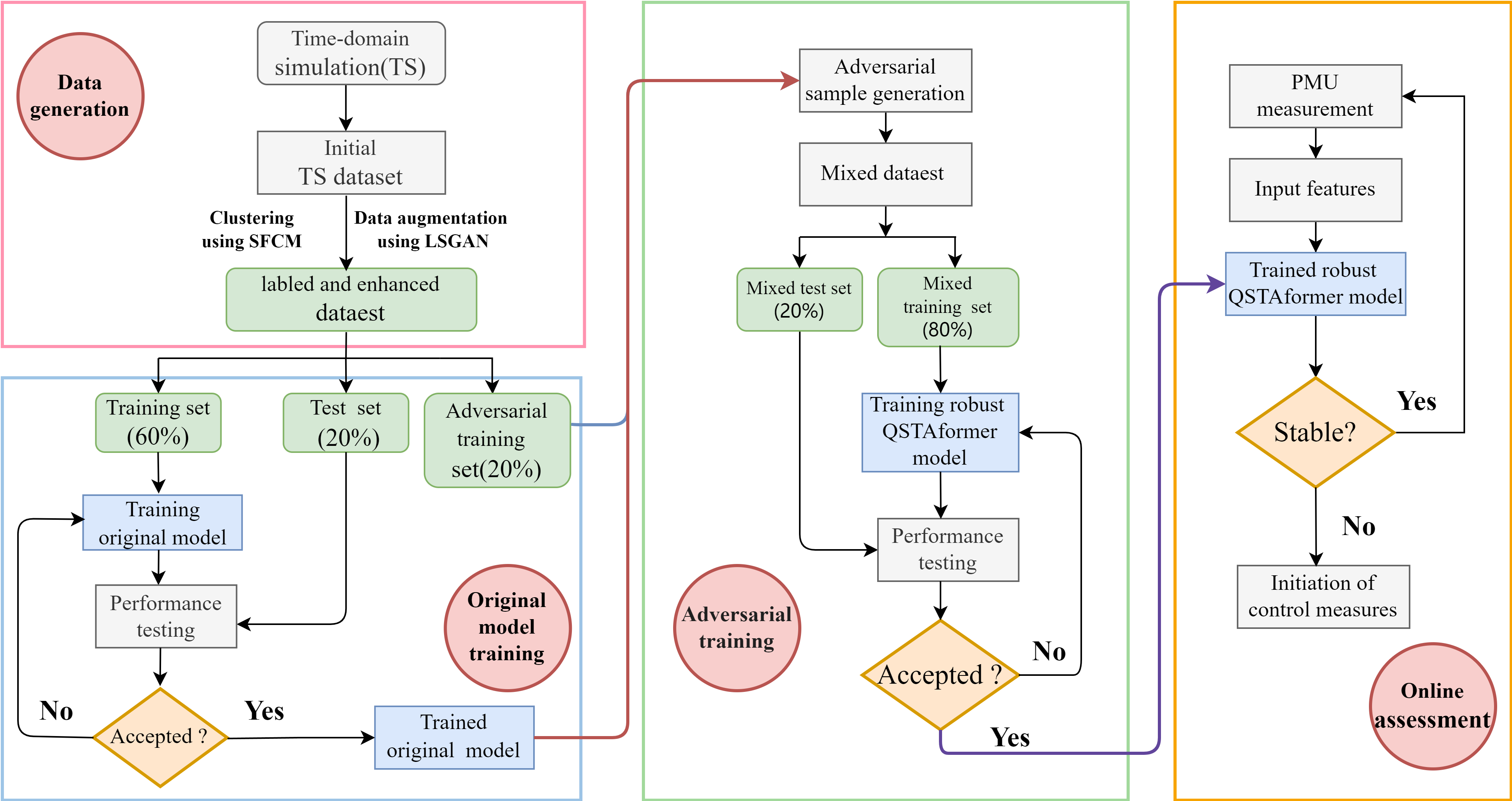}
    \caption{Overview of the proposed adversarially robust QML-based STVSA framework.}
    \label{fig:sys-eval}
\end{figure*}

To comprehensively evaluate short-term voltage stability under adversarial attacks, this paper proposes a QML-based real-time assessment framework, as illustrated in Fig. \ref{fig:sys-eval}. The workflow involves three major stages:
(i) original model training using labeled and augmented datasets generated via SFCM and LSGAN;
(ii) adversarial training through integration of crafted attack samples to improve robustness;
(iii) real-time evaluation under measurement inputs for practical deployment.
Each of these stages is elaborated in the following subsections.
\subsection{Original Model Training }
The initial dataset is generated using simulation software, followed by the application of multiple data processing methods to produce a final, usable clean dataset, which is then utilized for model training and other subsequent tasks.

\begin{figure}[htbp]
    \centering
    \includegraphics[width=\linewidth]{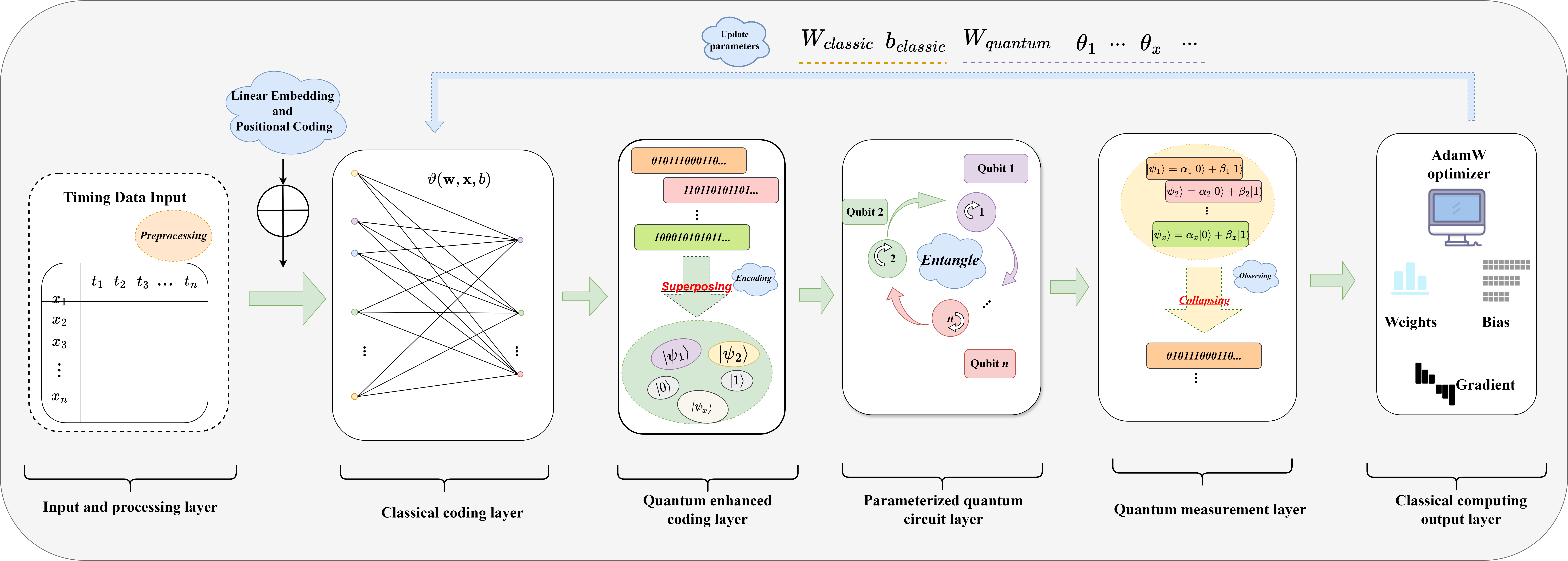}
    \caption{\textcolor{black}{The QSTAformer model architecture}}
    \label{STAformer}
\end{figure}

The general structure of the \textcolor{black}{QSTAformer} model proposed for training in this paper is illustrated in Fig. \ref{STAformer}.
\sloppy
{\color{black}
\begin{enumerate}
    \item As a data input layer, it accepts time series data input and preprocesses the data to make it suitable for the input form of the Transformer model.
    \item  The input data are first linearly embedded and positionally encoded, and are subsequently fed into the computational part of the classical model. 
    \item After processing in the first \emph{N}-1 layers, it is input to the quantum coding layer via the quantum residual projection layer, where the classical data is encoded into a quantum state.
    \item Quantum state data is fed into the PQCs and computed using pre-constructed quantum circuits with variable parameters continuously updated.
    \item Quantum state data is collapsed to classical data by quantum measurement, which is the process of conversion of quantum state to classical data.
    \item The classical data obtained after training is re-iterated for modeling based on feedback such as loss function and finally the best model obtained is saved.
\end{enumerate}

In the evaluation process, the model is based on the classical model, and the hybrid model makes full use of the inherent properties and laws of quantum computing, breaks through the secondary complexity bottleneck of the classical attention mechanism, and utilizes the high-dimensional Hilbert space with entanglement, which enhances the expression capability and realizes the parallel computation of data in the high-dimensional space. By integrating quantum complexity theory, entanglement representation and hybrid architecture design, the efficiency and expression boundaries of sequence modeling are redefined.}

\subsection{Adversarial Training}
\subsubsection{Adversarial Sample Generation}
This paper investigates three potent attack methodologies—MI-FGSM, PGD, and C\&W Attack—across two distinct scenarios: white-box and gray-box. \textcolor{black}{The different kinds of attacks in different scenarios and the intensity of the attacks reflect the different levels of knowledge of the attacker about the content information of the attack model.} Utilizing either part of the clean data or alternative models helps generate adversarial samples. By adjusting the perturbation strengths, adversarial samples with varying degrees of tampering are obtained.

\subsubsection{Model Robust Training}
As an efficient and practical method to resist adversarial attacks, adversarial training demonstrates excellent performance in improving model robustness \cite{25}. In this paper, the obtained attack samples are mixed with clean samples in a specified proportion, and the mixed dataset is then split into a training set and a test set.
The hybrid training set serves as input for continuous training of the original model. The \textcolor{black}{QSTAformer} model re-extracts data features, learns adversarial data characteristics, updates model weights and parameters, and redefines the classification boundaries. An independent hybrid data test set verifies model performance, ensuring a balance between accuracy and robustness.

\subsection{Real-Time Evaluation}
Deploying the robust model to counter this attack in the power system allows PMUs to gather essential power system characteristic quantities in real time. These inputs feed into the evaluation model to monitor the system's stability effectively.
If the system is stable, the monitoring system should continue to monitor it, if it happens that the system 
is categorized as unstable, the system immediately initiates contingency measures to 
ensure that the system does not experience a more serious incident. 

\section{Case Study}
\begin{figure}[H]
    \centering
    \includegraphics[width=1\linewidth]{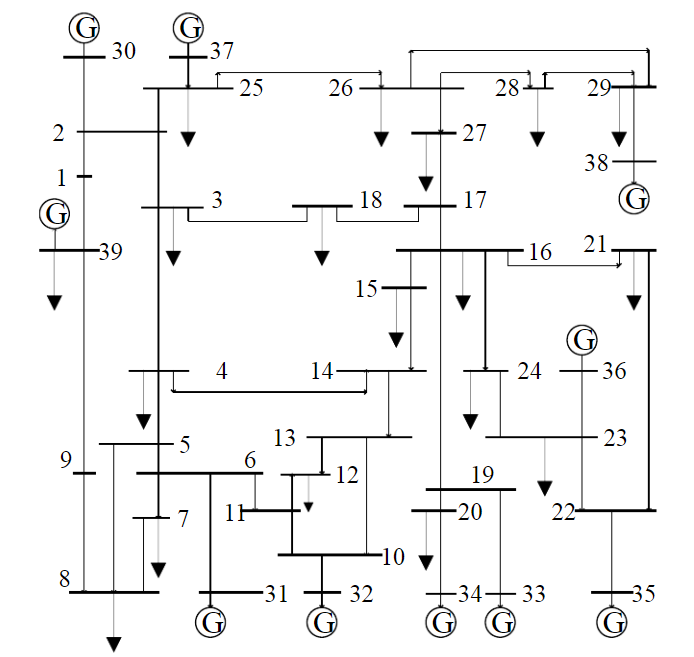}
    \caption{IEEE 39-bus test system diagram. }
    \label{fig5}
\end{figure}

To assess the performance of the proposed methodology, simulation tests were performed on the IEEE 39-bus power system, as shown in Fig. \ref{fig5}. The IEEE 39-bus test system consists of 39 buses, 10 generators, and 46 transmission lines, widely used in power system stability studies \cite{li2022deep, zhou2022noise, li2023pmu, zong2023transient, 12, li2025ai, bi2025power}.
 The experiments were conducted on a desktop computer with the following configuration: Intel Core i5-12600KF 3.70 GHz CPU, 32 GB RAM, and an RTX 4060 Ti GPU. All experiments presented in this paper were implemented using Python version 3.10, with PyTorch version 2.0.1+cu118. The coding and manipulation of quantum circuits were conducted using Pennylane, version 0.36.0, with quantum simulations executed on the default backend. 
 
{\color{black} Among the selected comparison methods, the quantum-enhanced LSTM (QLSTM) model serves as a paradigm for a quantum-classical hybrid model that facilitates the exploration of the quantum properties of the proposed method. The classical Transformer model, on the other hand, is the core architecture used in current general-purpose large language models and has shown good performance in many domains. Equally important, the classical LSTM structure lays the theoretical foundation of sequence modeling and has unique advantages in sequence data processing.}

Table \ref{tab:hyperparameters} summarizes the primary hyperparameters for SFCM and LSGAN, where $m$ denotes the index of the affiliation matrix, $\lambda$ is the regularization parameter for equilibrium, and $k$ controls the balance between the discriminator and generator. The hyperparameters for QSTAformer are provided separately in Table \ref{tab:qlstm}. \textcolor{black}{In addition, the hyperparameter tuning selection is performed in this paper with empirical as the main scope.}
\begin{table}[ht]
\centering
\caption{Selected Hyperparameters for the SFCM and the LSGAN}
\setlength{\tabcolsep}{20pt} 
\begin{tabularx}{0.48\textwidth}{@{} >{\raggedright\arraybackslash}X >{\raggedright\arraybackslash}X >{\raggedright\arraybackslash}X @{}}
\toprule
\textbf{Item} & \textbf{Parameters} & \textbf{Values} \\
\midrule
\multirow{4}{*}{SFCM} & $m$ & 2 \\
                      & $\lambda$ & 5 \\
                      & Max iterations & 1000 \\
                      
\midrule
\multirow{5}{*}{LSGAN} & Optimizer & Adam \\
                       & Learning rate & 1e-4 \\
                       & Batch size & 32 \\
                       & Epoch & 1000 \\
                       & Max iterations & 3000 \\
                       & $k$ & 5 \\
\bottomrule
\end{tabularx}
\label{tab:hyperparameters}
\end{table}
\begin{table}[H]
\centering
\caption{Hyperparameter settings of the QSTAformer model}
\fontsize{8pt}{10pt}\selectfont 
\setlength\tabcolsep{20pt} 
\begin{tabularx}{0.48\textwidth}{@{} l >
{\raggedright\arraybackslash}X l @{}}
\toprule
\textbf{Item} & \textbf{Parameters} & \textbf{Values} \\
\midrule
\multirow{9}{*}{QSTAformer} & Quantum bit number & \textcolor{black}{4} \\
                       & Quantum layer number & \textcolor{black}{4} \\
                       & Coding mode& Angle coding \\
                       & Entangled mode& \textcolor{black}{Ring entanglement} \\
                       & Optimizer & \textcolor{black}{AdamW} \\
                       & Learning rate range & \textcolor{black}{1e-4 - 1e-5} \\
                       & Batch size & \textcolor{black}{32} \\
                       & Epoch & \textcolor{black}{55}\\
\bottomrule
\end{tabularx}
\label{tab:qlstm}
\end{table}

\subsection{Dataset Generation}
A total of 2040 original samples were collected using the commercial simulation software PSD-BPA, comprising 192 stable, 698 unstable, and 1150 undecidable cases. The samples were generated under varying simulation conditions, including different load levels, induction motor ratios, and fault scenarios. The electrical quantities that reflect  the real-time state of the system—active power, reactive power, and voltage magnitude (denoted as $P$, $Q$ and $U$)—were selected to construct the time-series dataset. The detailed simulation conditions used in dataset generation are summarized in Table~\ref{tab:simulation_conditions}.

\begin{table}[htbp]
\centering
\caption{Simulation conditions for dataset generation}
\fontsize{8pt}{10pt}\selectfont %
\setlength{\tabcolsep}{15pt} %
\begin{tabularx}{0.48\textwidth}{@{} l >{\raggedright\arraybackslash}X @{}}
\toprule
\textbf{Sets} & \textbf{Values} \\
\midrule
Load level & 80\%, 90\%, 100\%, 110\%, 120\% \\
Induction motor load ratios & 70\%, 80\%, 90\% \\
Fault location & 0\%, 25\%, 50\%, 75\% \\
Fault type & Three-phase short circuit \\
Fault clearing time (s) & 0.05 (near end), 0.1 (far end) \\
\bottomrule
\end{tabularx}
\label{tab:simulation_conditions}
\end{table}

To initiate the labeling process for semi-supervised clustering, a small subset of simulation samples exhibiting clearly distinguishable voltage behavior was heuristically labeled based on domain knowledge. For the remaining unlabeled samples with ambiguous characteristics, the SFCM algorithm was applied to infer their labels using the partial prior knowledge. This procedure yielded a dataset containing 979 stable and 1061 unstable samples.

To further expand the dataset and improve class balance, the LSGAN algorithm was employed to generate synthetic samples. The final augmented dataset comprises 10,000 instances, with 5,162 labeled as stable and 4,838 as unstable. 

\subsection{Quantitative Validation of LSGAN Samples}

To guarantee that the samples generated by LSGAN are statistically consistent with real PMU measurements and suitable for downstream stability classification, a two-level validation strategy is adopted. First, distributional similarity is examined using Maximum Mean Discrepancy (MMD) with an RBF kernel. The obtained value, $\mathrm{MMD}=0.00386$, is extremely small, indicating that the synthetic sequences closely match the underlying temporal distribution of the real PMU voltage trajectories.

To further assess functional reliability, we employ Train-on-Real--Test-on-Synthetic (TSTR) and Train-on-Synthetic--Test-on-Real (TRTS) evaluations. Both cross-domain results deviate by less than 5\% from the Real--Real baseline in terms of Accuracy, AUC, and F1-score. This demonstrates that the generated samples not only preserve the statistical structure of original simulation-based PMU data but also support consistent task-level performance without introducing noticeable distributional bias.

Overall, these results confirm that the LSGAN-based augmentation maintains the temporal and statistical characteristics required for voltage stability assessment and enhances sample diversity in a controlled and reliable manner.

\textcolor{black}{\subsection{Classification Performance Comparison}}
In this study, we used classical binary classification metrics, accuracy, F1-score, and Area Under the Curve (AUC) to provide a comprehensive evaluation of the classification performance of the proposed methods, along with analyses of different arithmetic examples based on rating metrics from various models\cite{luo2021data}.
\textcolor{black}{Note that in this study, all classification metrics (Accuracy, F1-score, and AUC) are normalized to a [0, 1] scale in both text and figures for consistency and interpretability.}

\begin{figure}[htbp]
    \centering
    \includegraphics[width=1\linewidth]{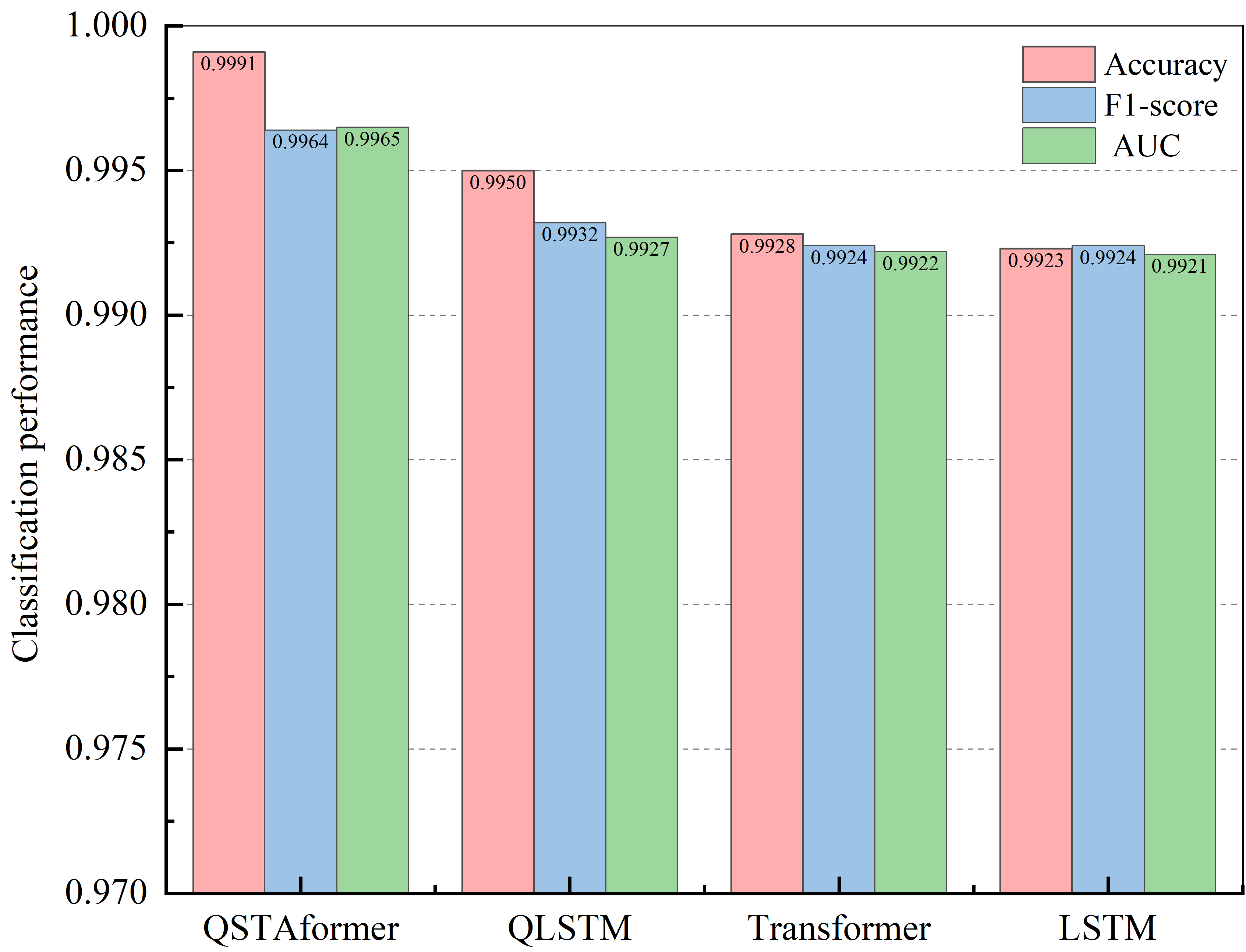}
    \caption{\textcolor{black}{Comparison of classification performance of different models}}
    \label{model}
\end{figure}
{\color{black}
Based on the analysis of the graphs depicted  in Fig. \ref{model}, it shows that the QSTAformer model achieves Accuracy 0.9985, F1-score 0.9979, and AUC 0.9972 in the STVSA, which is better than the comparison examples. The performance advantage stems from the quantum-classical double gain: compared with the classical transformer, the QSTAformer replaces the traditional self-attention with an 8-qubit parameterized circuit at the last layer, embeds the features into the high-dimensional Hilbert space, and utilizes the quantum entanglement and superposition to capture the nonlinear small perturbations during the critical voltage destabilization, so as to realize the gradient of the voltage stability. The potential acceleration of gradient complexity from $\Omega(T^{2})$ to $\Omega(T)$ significantly improves the convergence speed and robustness, compared with QLSTM, the QSTAformer preserves Transformer's multihead self-attention mechanism, which avoids gradient decay caused by cyclic structure while dealing with the long-range spatiotemporal dependence in parallel, leading to higher batch training efficiency and smaller error propagation. This hybrid architecture can be deployed in Noisy Intermediate-Scale Quantum (NISQ) hardware with only a single layer of quantum circuits, providing a highly accurate, low false alarm, and grounded quantum-classical collaborative solution for online stability assessment of power grids.}

{\color{black}  
\subsection{Convergence Performance Comparison}

\begin{figure}
    \centering
    \includegraphics[width=1\linewidth]{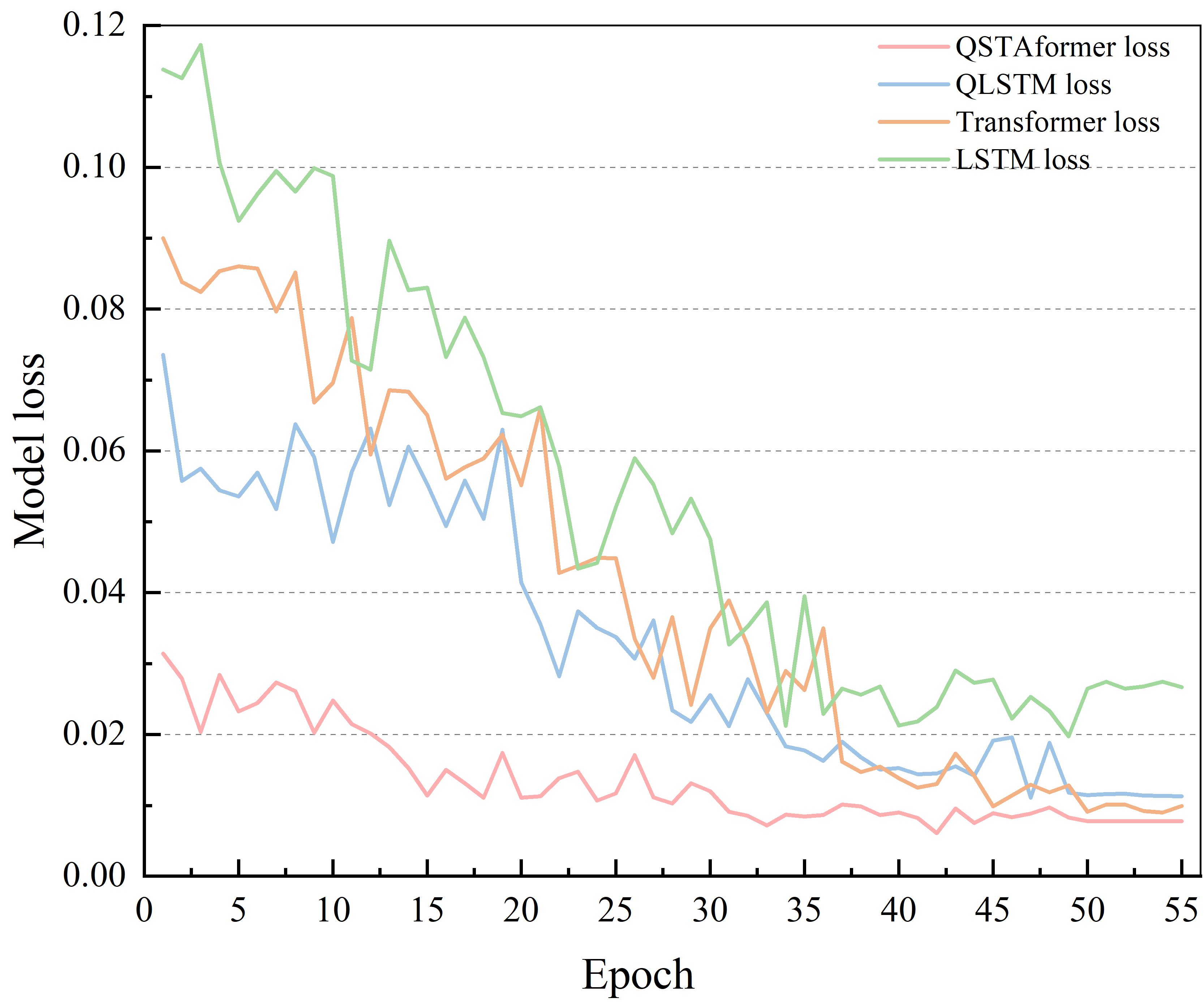}
    \caption{\textcolor{black}{Comparison of convergence of different models}}
    \label{convergence}
\end{figure}
As illustrated in Fig. \ref{convergence}, the QSTAformer model demonstrates the fastest convergence among the evaluated models, with the training loss dropping below 0.02 within approximately 13 epochs. It achieves the lowest final loss value and the steepest descent rate, indicating superior convergence performance compared to QLSTM, Transformer, and LSTM models. Although both QLSTM and Transformer also exhibit relatively good convergence, their loss reduction rates and final values are notably inferior to those of the QSTAformer. The LSTM model, in contrast, converges the slowest and shows the least favorable convergence behavior.

The enhanced convergence of the QSTAformer model can be attributed to the synergistic integration of its quantum-enhanced module and Transformer architecture. The quantum module, leveraging PQCs, introduces a high-dimensional quantum feature space, which enables the model to capture complex nonlinear patterns and long-range dependencies more effectively, thereby accelerating the loss minimization process. Simultaneously, the multi-head self-attention mechanism in the Transformer component strengthens the model’s capability to handle long sequential data, further improving training efficiency.

Such convergence characteristics hold significant practical relevance. First, the rapid convergence of the QSTAformer allows it to reach desirable performance in fewer iterations, thereby reducing computational cost and training time. Second, faster convergence reduces the risk of becoming trapped in local minima and contributes to better generalization performance. Lastly, for real-time applications such as voltage stability assessment, the prompt convergence of the QSTAformer facilitates timely and reliable predictions, which are critical for maintaining the safety and stability of power system operations.
In summary, the superior convergence performance of the QSTAformer model highlights the effectiveness of integrating quantum enhancement with Transformer-based mechanisms and underscores its practical value for energy system applications.}

{\color{black} 
\subsection{Performance Comparison of Different Qubits}
\begin{figure}[htbp]
    \centering
    \includegraphics[width=1\linewidth]{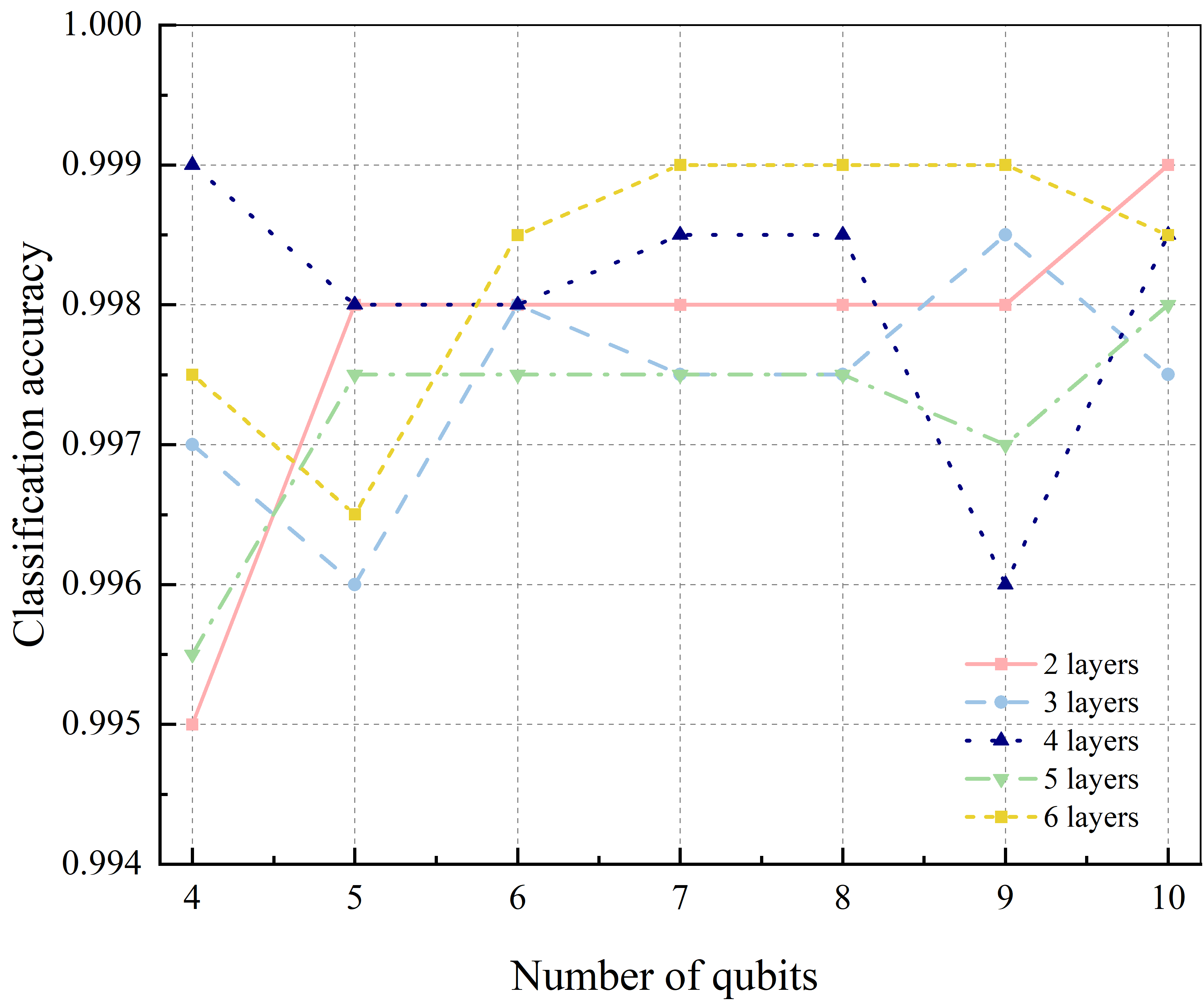}
    \caption{\textcolor{black}{Comparison of classification performance under different qubits}}
    \label{different_qubits.png}
\end{figure}

In QML, the number of quantum bits, as one of the key design factors, has a significant impact on model performance. 
As depicted  in Fig. \ref{different_qubits.png}, this paper analyzes the effect of variations in the number of quantum bits (4 to 10) and the number of layers (2 to 6) on the classification accuracy of the model.
The results indicate that in shallow models (2-layer circuits), increasing the number of qubits consistently improves performance, with accuracy rising from 0.9950 (4 qubits) to 0.9990 (10 qubits), suggesting that additional quantum resources enhance the model’s expressive capacity.

However, in deeper models (3 to 6 layers), the performance exhibits no clear trend and fluctuates between 0.9950 and 0.9990. For instance, in the 4-layer configuration, accuracy drops from 0.9990 (4 qubits) to 0.9965 (8 qubits), then recovers to 0.9985 (10 qubits), this inconsistency stems from the increased complexity of optimization in deeper architectures. Furthermore, the performance appears to saturate across all configurations, with accuracy consistently within the range of 0.9950 to 0.9990, indicating diminishing returns as model complexity increases. These findings suggest that the influence of qubit count is highly dependent on circuit depth—shallow models benefit more clearly from additional qubits, while deeper models require a careful balance between capacity and trainability.}

Although increasing the number of qubits improves the expressive power of parameterized quantum circuits, it also makes optimization substantially more difficult. As the Hilbert-space dimension grows exponentially, variational circuits tend to exhibit flatter gradients and even barren-plateau behavior, which suppresses effective parameter updates. Under the fixed 4-layer PQC structure used in our experiments, this effect becomes more noticeable as the qubit number increases from 4 to 8, resulting in a slight but consistent accuracy reduction (approximately 0.05\%). Because all experiments were executed on the PennyLane default noiseless backend, the performance degradation originates purely from optimization difficulty rather than quantum noise.

When the circuit depth increases, additional entanglement partially mitigates this expressivity–trainability trade-off, producing the non-monotonic accuracy fluctuations observed in Fig. \ref{different_qubits.png}. This phenomenon is well aligned with recent findings on the trainability of variational quantum models, providing a physically interpretable explanation for why performance improves steadily in shallow circuits but oscillates or saturates in deeper architectures.

{\color{black}
\subsection{Performance Comparison of Different Quantum Layers}
In QML, the depth of the quantum circuit also has an important impact on the performance of the model.
\begin{figure}[htbp]
    \centering
    \includegraphics[width=1\linewidth]{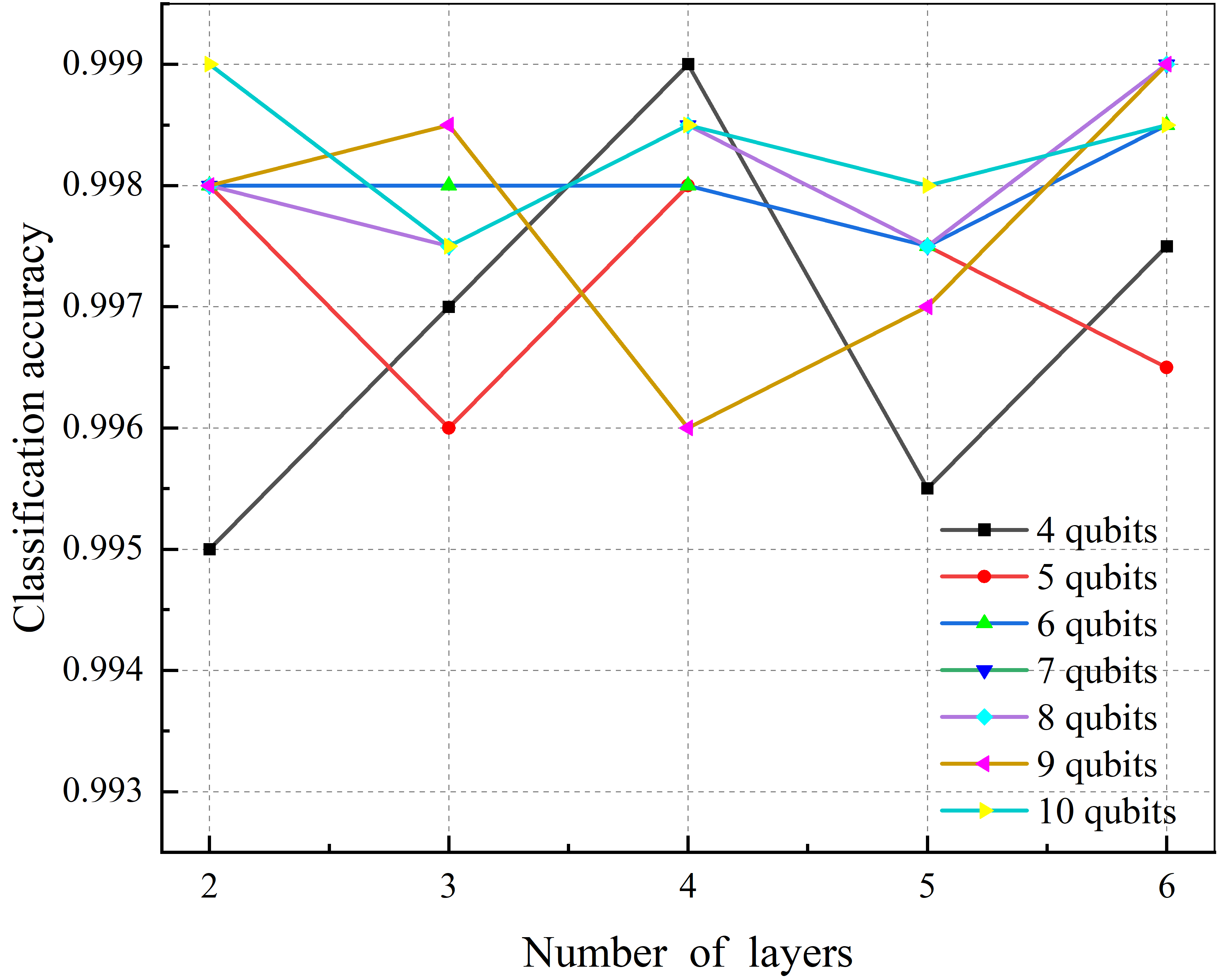}
    \caption{\textcolor{black}{Comparison of classification performance under various quantum layers}}
    \label{quantum_layers}
\end{figure}

As illustrated  in Fig. \ref{quantum_layers},  when the number of qubits is held constant, the model performance exhibits a nonlinear and non-monotonic trend with increasing quantum circuit depth. Overall, shallow quantum models (2–4 layers) demonstrate more stable and superior performance across most qubit configurations. For instance, with 4 qubits, the accuracy steadily increases from 0.9950 at 2 layers to 0.9990 at 4 layers; with 5 qubits, the accuracy remains consistently high at 0.9980 for both 2 and 4 layers. These results indicate that shallow quantum circuits are effective in feature representation and convergence behavior.

However, as the circuit depth increases further (5–6 layers), the performance tends to fluctuate or even degrade under several qubit settings. For example, in the case of 4 qubits, the model reaches a peak accuracy of 0.9990 at 4 layers but drops to 0.9955 at 5 layers, with a slight rebound at 6 layers that still falls short of the peak. Similar trends are observed for qubit counts of 6, 8, and 10. This performance inconsistency may be attributed to overfitting in deeper models or increased optimization complexity. Moreover, as the overall model accuracy is already near saturation (above 0.9950), further increases in circuit depth provide only marginal gains.

Taken together, these findings suggest that the effect of circuit depth on model accuracy is highly dependent on the number of qubits. Shallow circuits benefit more clearly from available quantum resources, whereas deeper circuits require a careful balance between model capacity and trainability. This highlights the importance of jointly optimizing circuit depth and qubit count in hybrid quantum neural networks to avoid over-parameterization and ensure generalization performance.
}

\subsection{ Performance Comparison of Different Sampling Times}
Based on the IEEE/CIGRE Joint Investigation Group on Stability Terms and Definitions Study 
\cite{28}, the time scale for studying STVS is typically in the order of several seconds. According to the $\text{IEEE Std C37.242}\text{-2021}$, the typical communication delay time for PMU data acquisition systems ranges from 0.02 to 0.05 seconds \cite{9665413}. 
Furthermore, given the response time requirements for emergency measures following instability, 
the online real-time assessment model must quickly determine the system's state within a very short period. 
Therefore, this paper investigates the impact of varying time scales on the QSTAformer model's performance by employing different sampling times.
{\color{black}
As shown in Fig. \ref{sampling_times.png}, the model performance exhibits a nonlinear dynamic evolution as the sampling time window increases. Experimental results indicate that within the interval of 0.03 to 0.09 seconds, the model undergoes a phase of rapid performance improvement, reaching a peak accuracy of 0.9990 at 0.09 seconds. Beyond this point, from 0.09 to 0.2 seconds, the performance tends to saturate and fluctuates slightly around the peak value. The loss curve mirrors this trend, displaying a sharp decline followed by stable convergence, indicating that the model has been effectively trained without signs of overfitting.

\begin{figure}[htbp]
    \centering
    \includegraphics[width=1\linewidth]{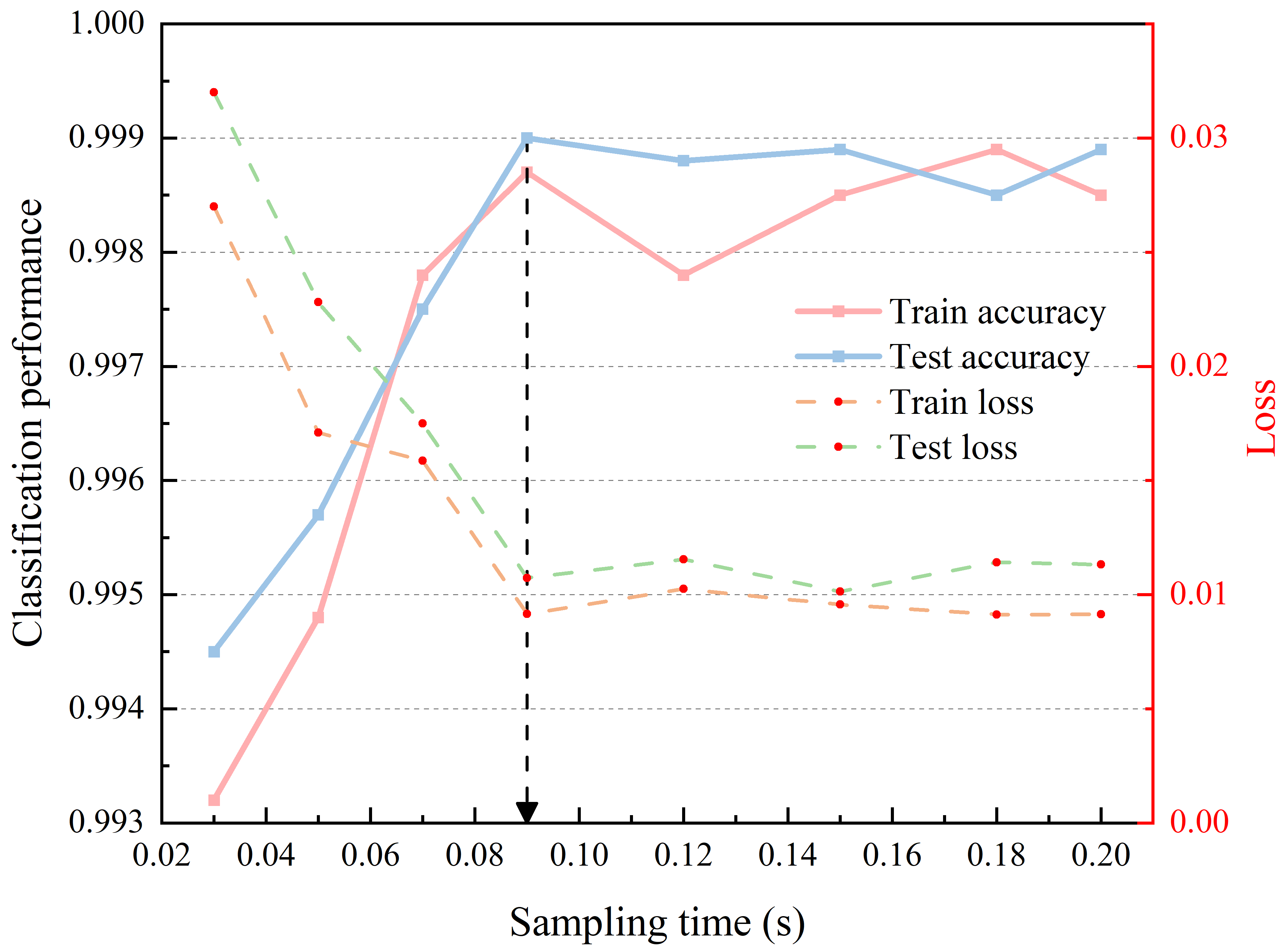}
    \caption{\textcolor{black}{Comparison of classification performance under different sampling times}}
    \label{sampling_times.png}
\end{figure}
This phenomenon reveals a complex coupling between the efficiency of feature extraction and the system’s response time. Notably, while extending the sampling window improves data resolution, it also introduces redundant information that leads to performance saturation. Therefore, 0.09 seconds is identified as the optimal sampling time window in this study, as it achieves a favorable balance between real-time system responsiveness and classification accuracy.}

\subsection{Ablation Study on the Hybrid Pipeline}

To rigorously verify the effectiveness of the proposed hybrid framework, a systematic ablation analysis is conducted, examining the contribution of each module and their combined effect. As shown in Table~\ref{tab:ablation-config}, four configurations are evaluated: (i) the QSTAformer-only baseline, (ii) QSTAformer with SFCM-based label refinement, (iii) QSTAformer with LSGAN-based data augmentation, and (iv) the complete hybrid pipeline incorporating both components. Throughout this study, the QSTAformer-only model serves as the reference baseline.

\begin{table}[t]
\centering
\caption{Ablation configurations of the hybrid pipeline}
\begin{tabular}{p{3.1cm} p{4.8cm}}
\toprule
\textbf{Configuration} & \textbf{Description} \\
\midrule
QSTAformer-only & Baseline model with no preprocessing or augmentation \\
+SFCM & Adds sample--feature clustering for label refinement and noise mitigation \\
+LSGAN & Introduces GAN-based data augmentation to enhance data diversity and alleviate class imbalance \\
Full Hybrid & Integrates both SFCM and LSGAN before QSTAformer training \\
\bottomrule
\end{tabular}
\label{tab:ablation-config}
\end{table}

To comprehensively measure the effect of each component, we report the clean accuracy, average robust accuracy (averaged across MI-FGSM, PGD, and C\&W attacks), and robustness drop (defined as the difference between clean and robust accuracy). The experimental results are presented in Table~\ref{tab:ablation-results}.

\begin{table}[t]
\centering
\caption{Performance comparison under different ablation configurations}
\label{tab:ablation}
\resizebox{\linewidth}{!}{
\begin{tabular}{lccc}
\hline
\textbf{Model} & \textbf{Clean Acc (\%)} & \textbf{Avg. Robust Acc (\%)} & \textbf{Robustness Drop (\%)} \\
\hline
QSTAformer-only & 63.89 & 54.17 & 9.72 \\
+SFCM           & 79.78 & 73.17 & 6.61 \\
+LSGAN          & 88.50 & 82.25 & 6.25 \\
Full Hybrid     & 99.91 & 95.43 & 4.48 \\
\hline
\end{tabular}
}
\label{tab:ablation-results}
\end{table}

The results reveal a clear performance improvement trend as additional components are incorporated. SFCM significantly enhances label reliability by reducing label noise and improving sample consistency, leading to simultaneous gains in both clean and robust accuracy. LSGAN further boosts performance by generating diverse and balanced data samples, which improves generalization under both clean and perturbed inputs. When combined, the full hybrid configuration achieves the highest clean accuracy (99.91\%) and the smallest robustness drop (4.48\%), demonstrating that label refinement and augmentation provide complementary benefits.

These findings remain consistent under multiple random seeds, with performance fluctuations below 1\% for both clean and adversarial metrics, confirming the statistical stability of the hybrid pipeline. Overall, the ablation study validates that the hybrid architecture substantially strengthens generalization and robustness, and is therefore a necessary component of the overall model design.

\subsection{Adversarial Conditions and Physical Interpretation}

To ensure that adversarial evaluation is physically meaningful and reproducible, we explicitly define the adversarial conditions used throughout the experiments. Two threat scenarios are considered: a white-box adversary with full gradient access, representing a fully compromised interface, and a gray-box adversary with limited knowledge, representing realistic PMU-level or communication-level intrusions.

Three representative attack methods are implemented, namely MI-FGSM and PGD under the $\ell_{\infty}$-norm and the C\&W attack under the $\ell_{2}$-norm. The perturbation strength is controlled by $\varepsilon \in \{0.01, 0.03, 0.05\}$ p.u., and the C\&W parameters are set with confidence $c=2$, learning rate $0.05$, and a maximum of $100$ iterations. All attacks are implemented using custom PyTorch modules to guarantee deterministic behavior.

Adversarial perturbations are injected into the PMU voltage-magnitude channels before entering the model, reflecting realistic scenarios such as measurement interference, communication distortion, or malicious data manipulation in wide-area monitoring systems. Notably, the perturbation levels used in this study fall within the typical PMU noise and error envelope specified in IEEE C37.118, ensuring that the adversarial conditions remain physically plausible and system-relevant.

For each perturbation level, we report the clean accuracy, adversarial accuracy, and robustness drop under identical training settings to isolate the impact of attacks. All results in this subsection are produced under this unified and fully reproducible evaluation protocol.

\subsection{Robustness Analysis Against Attacks }
\subsubsection{Vulnerability of The Original Model}
\begin{figure}[htbp]
    \centering
    \includegraphics[width=1\linewidth]{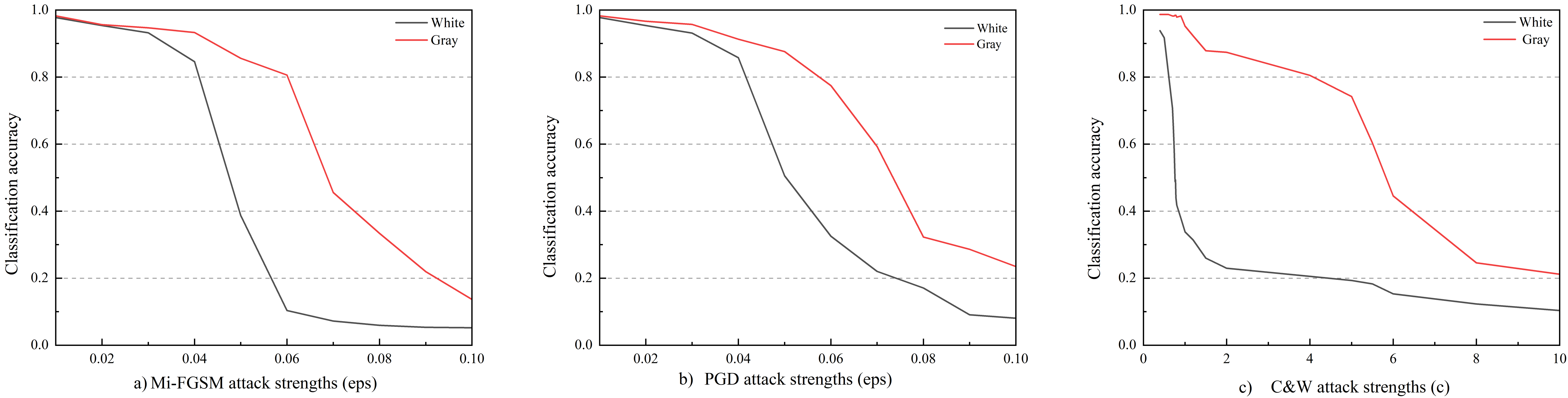}
    \caption{\textcolor{black}{Accuracy of the original model after attack}}
    \label{attack}
\end{figure}

As depicted in Fig. \ref{attack}, this study reveals the variation patterns of the original model’s evaluation accuracy under white-box and gray-box scenarios through three typical adversarial attack methods: as the attack strength or magnitude of the disturbance progressively increases, the model accuracy in both scenarios exhibits a significant downward trend. Notably, while the gray-box scenario demonstrates relatively stronger perturbation resistance (tolerating larger perturbation magnitudes than the white-box scenario), the model performance still degrades sharply with escalating attack intensity. 
Specifically, under the gray-box scenario, the model evaluation accuracy for the C\&W attack stabilizes around 20\% even when the perturbation magnitude continues to increase, displaying distinct performance convergence characteristics.
In contrast, MI-FGSM and PGD attacks, which utilize gradient-based update mechanisms, are more destructive: As the attack strength intensifies, the model performance ultimately collapses completely, resulting in systematic misclassifications by the evaluation model. These adversarial attack experiments conclusively validate the inherent vulnerabilities of data-driven models in security defense mechanisms.

\subsubsection{Robustness After Adversarial Training}
\begin{figure}[htbp]
    \centering
    \includegraphics[width=1\linewidth]{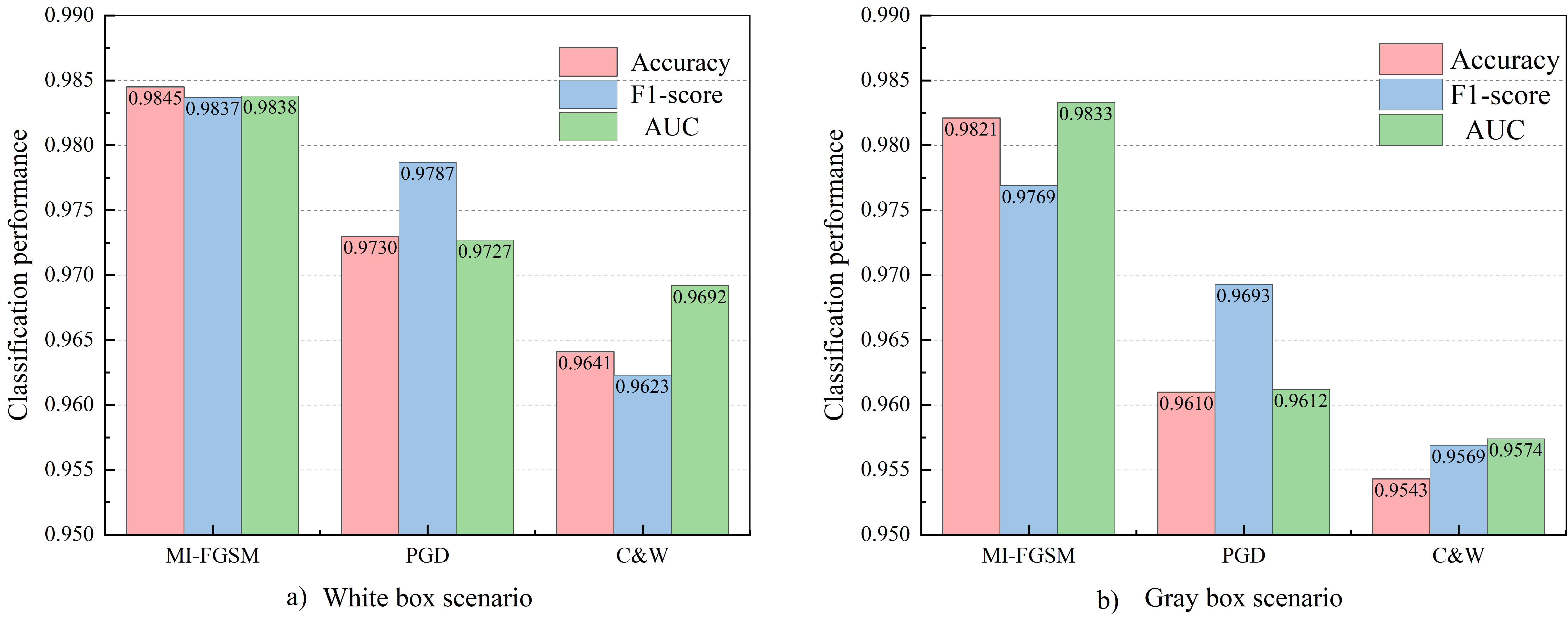}
    \caption{\textcolor{black}{Model robustness against MI-FGSM, PGD, and C\&W Attacks in White-Box and Gray-Box scenarios}}
    \label{robustness}
\end{figure}

As depicted  in Fig. \ref{robustness}, the original model, after being trained with  adversarial samples, demonstrates prominent robustness against both clean and adversarial samples. 
Experimental results demonstrate that the model maintains a stable decision boundary across both pristine samples and adversarial samples in the testing sets. In particular, while adversarial training induces a controlled reduction in absolute recognition accuracy, overall evaluation accuracy consistently adheres to industrial-grade application standards ($\geqslant95\%$). This technical approach achieves significantly improved adversarial robustness with an accuracy loss that does not exceed 5\%, providing a novel solution to build secure and reliable intelligent evaluation systems. 

{\color{black}
\section{Conclusions}
To address the limitations of classical machine learning models in STVSA under adversarial attacks, this study proposes a QSTAformer architecture that integrates PQCs into the attention mechanism and incorporates a customized adversarial training strategy to enhance robustness. Comprehensive experiments validate the effectiveness, efficiency, and robustness of the proposed framework.}
The main conclusions of this study are summarized as follows: 
{\color{black}
\begin{enumerate}
    \item  The QSTAformer model demonstrates superior classification performance, achieving an accuracy of up to 0.9990 while maintaining fast convergence and low parameter complexity. This confirms the advantage of quantum-classical hybrid modeling for STVSA tasks.
    \item Systematic benchmarking of diverse PQC structures reveals their influence on convergence behavior, expressiveness, and inference efficiency, offering design insights for scalable quantum architectures in power system applications. 
    \item Robustness evaluations under white-box and gray-box adversarial attacks show that the adversarially trained QSTAformer model maintains classification accuracy above 0.9543 under strong perturbations, significantly outperforming classical baselines in security-sensitive scenarios.
    \item Case studies on the IEEE 39-bus system demonstrate the model’s practical potential in achieving secure, scalable, and generalizable voltage stability assessment under adversarial attacks.
\end{enumerate}}

This study is based on idealized quantum circuit simulations and does not consider hardware-level constraints such as noise, decoherence, and limited qubit connectivity. Future work will focus on implementing QML models on real quantum devices, exploring diverse quantum neural network architectures, and developing adaptive defense strategies to address evolving cyber-physical threats in power systems. {\color{black} Another interesting topic is to investigate privacy-preserving federated training and topology-aware architectures for QML models, as inspired by prior work \cite{li2022detection, ren2024esqfl, qu2024localization}.} 

\section*{Acknowledgements}
This work is supported by the Natural Science Foundation of China under Grant No. 52377081.

\bibliographystyle{IEEEtran}
\small\bibliography{Apen_repaire.bbl}
\end{document}